\documentclass[preprint2]{aastex}

\newcommand\plotthree[3]{\centering \leavevmode
 \columnwidth=.99\columnwidth 
 \includegraphics[width={\columnwidth}]{#1}%
 \hfil 
 \includegraphics[width={\columnwidth}]{#2}%
 \hfil 
 \includegraphics[width={1.4\columnwidth}]{#3}%
}%

\def\deg{$^{\circ}$}

\shorttitle{Bubbles in M87}
\shortauthors{Churazov et al.}

\begin{document}

\title{Evolution of Buoyant Bubbles in M87}

\author{E.Churazov\altaffilmark{1,2}, M.Br\"{u}ggen\altaffilmark{1,3},
C.R.Kaiser\altaffilmark{1}, H.B\"{o}hringer\altaffilmark{4},
W.Forman\altaffilmark{5}} 

\altaffiltext{1}{MPI f\"{u}r Astrophysik, Karl-Schwarzschild-Strasse 1, 85740
Garching, Germany}
\altaffiltext{2}{Space Research Institute (IKI), Profsoyuznaya 84/32,
Moscow 117810, Russia} 
\altaffiltext{3}{Churchill College, Storey's Way, Cambridge CB3 0DS, United
Kingdom} 
\altaffiltext{4}{MPI f\"{u}r Extraterrestrische Physik, P.O.Box 1603, 85740
Garching, Germany} 
\altaffiltext{5}{Harvard-Smithsonian Center for Astrophysics, 60 Garden St.,
Cambridge, MA 02138}

\begin{abstract}
The morphology of the X--ray and radio emitting features in the
central $\sim$ 50 kpc region around the galaxy M87 strongly suggests
that buoyant bubbles of cosmic rays (inflated by an earlier nuclear
active phase of the galaxy) rise through the cooling gas at roughly
half the sound speed. In the absence of strong surface tension,
initially spherical bubbles will transform into tori as they rise
through an external medium. Such structures can be identified in the
radio images of the halo of M87. During their rise, bubbles will
uplift relatively cool X--ray emitting  gas from the central regions
of the cooling flow to larger distances. This gas is colder than the
ambient gas and has a higher volume emissivity. As a result, rising
``radio'' bubbles may be trailed by elongated X--ray features as
indeed is observed in M87. We performed simple hydrodynamic
simulations to qualitatively illustrate the evolution of buoyant
bubbles in the M87 environment.

\end{abstract}


\label{firstpage}

\sloppypar

\section{Introduction}
The giant elliptical galaxy M87 continues to be the subject of
numerous experimental and theoretical studies. It is located at, or
near, the center of the very nearby ($\sim$ 18 Mpc, 1 arcminute
corresponds to $\sim$5 kpc) irregular Virgo cluster and is at the
center of a weak cooling flow with an estimated mass deposition rate
of some 10-40 $M_\odot$ per year (e.g. Peres et al. 1998). The radio
source Virgo A (3C274), associated with the galaxy M87, is well known
because of the spectacular jet which is observed both in
radio and optical bands (see e.g. Biretta 1999 for a review). In
addition to the jet, which is contained within the central 2 kpc or
$30''$ region, there is a lower surface brightness radio halo
extending up to a distance of $\sim$40--50 kpc from the nucleus
(e.g. Kassim et al. 1993, B{\"o}hringer et al. 1995, Owen, Eilek and
Kassim 1999, 2000).

The X--ray emission is strongly peaked at the position of the nucleus
of M87 and is largely symmetric around it. However, it was found in
the Einstein HRI observations (Feigelson et al. 1987) and then in the
ROSAT data (B{\"o}hringer at al. 1995) that there are departures from
a symmetric model and these departures resemble the morphology of 
some prominent features in the outer radio lobes. These findings initiated
numerous discussions on the interaction of the radio and X--ray
emitting plasmas (e.g. Feigelson et al. 1987, B{\"o}hringer et
al. 1995, B{\"o}hringer et al. 1999, Owen, Eilek and Kassim 1999,2000,
Harris et al. 1999). In a number of other cooling flow clusters with
strong radio halos surrounding the dominant galaxy (e.g. Perseus --
B{\"o}hringer at al. 1993, McNamara et al. 1996, Churazov et al. 2000,
Fabian et al. 2000a, A4059 -- Huang and Sarazin 1998, Hydra A --
McNamara et al. 2000) there is a clear anticorrelation of 
the radio and X--ray emitting plasmas. This makes the case of M87
especially interesting.

We discuss below one particular model which is able to reproduce
qualitatively both the radio and X--ray morphologies observed in
the central region surrounding the M87 galaxy.

\begin{figure*}
\plotthree{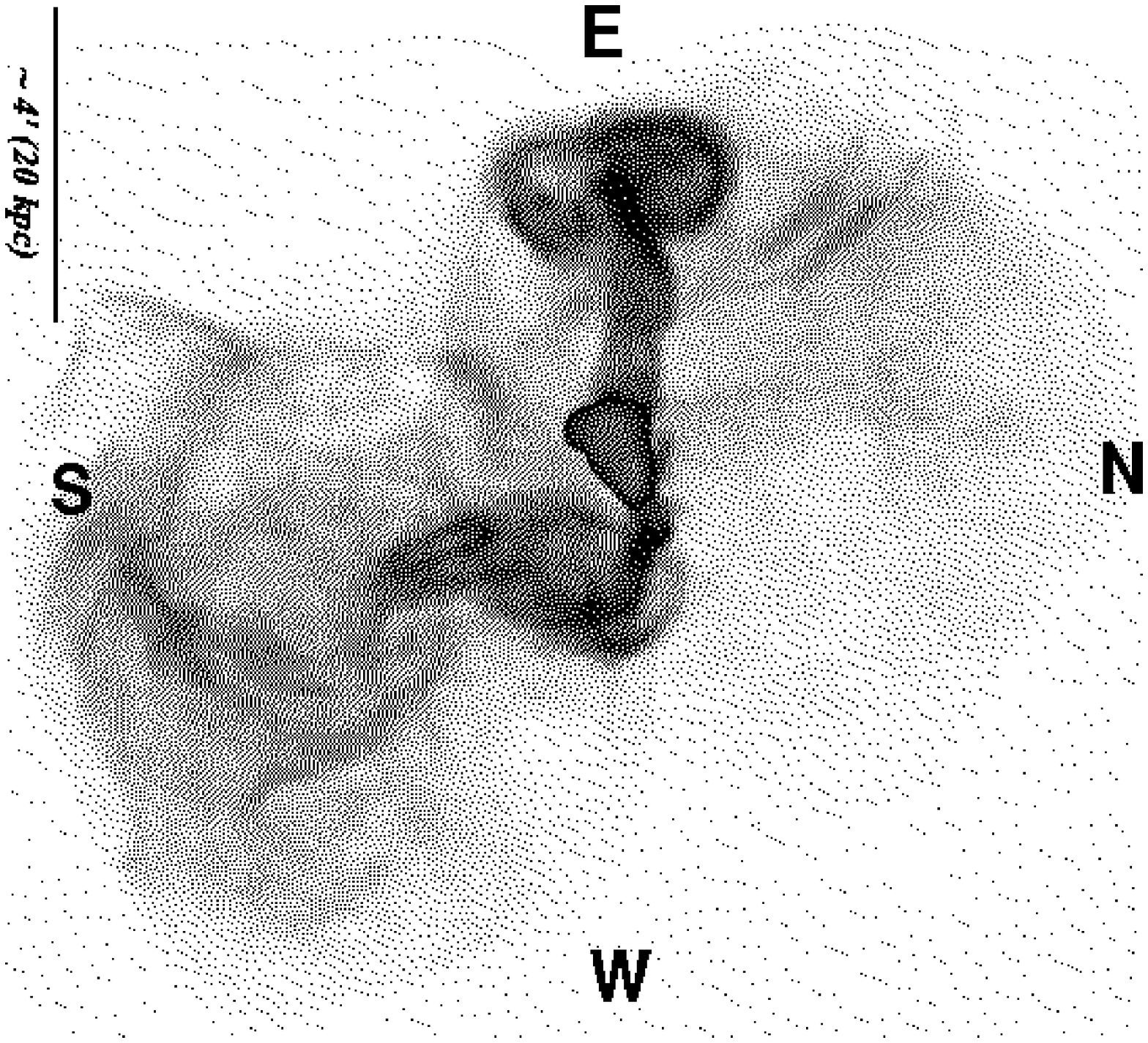}{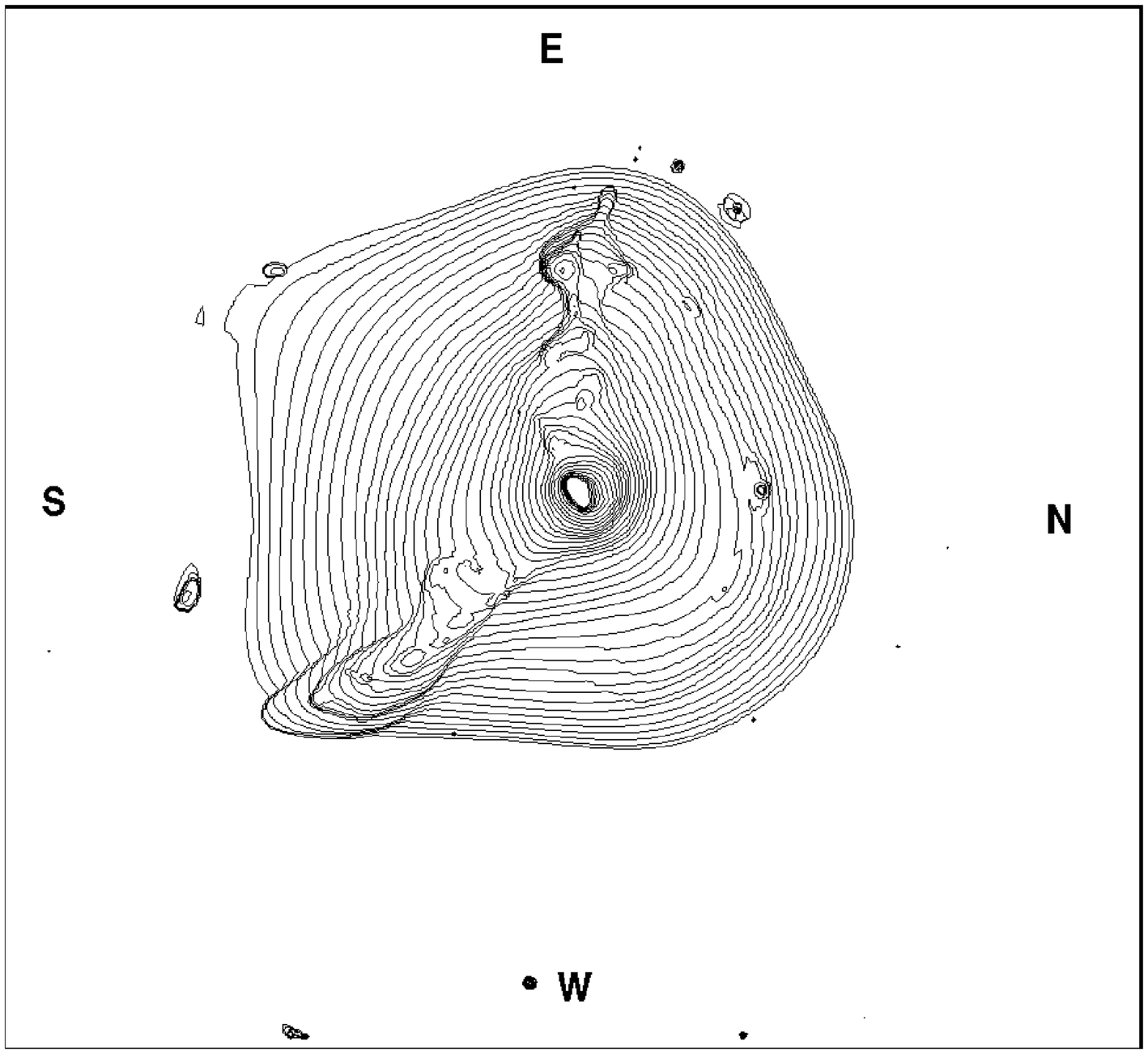}{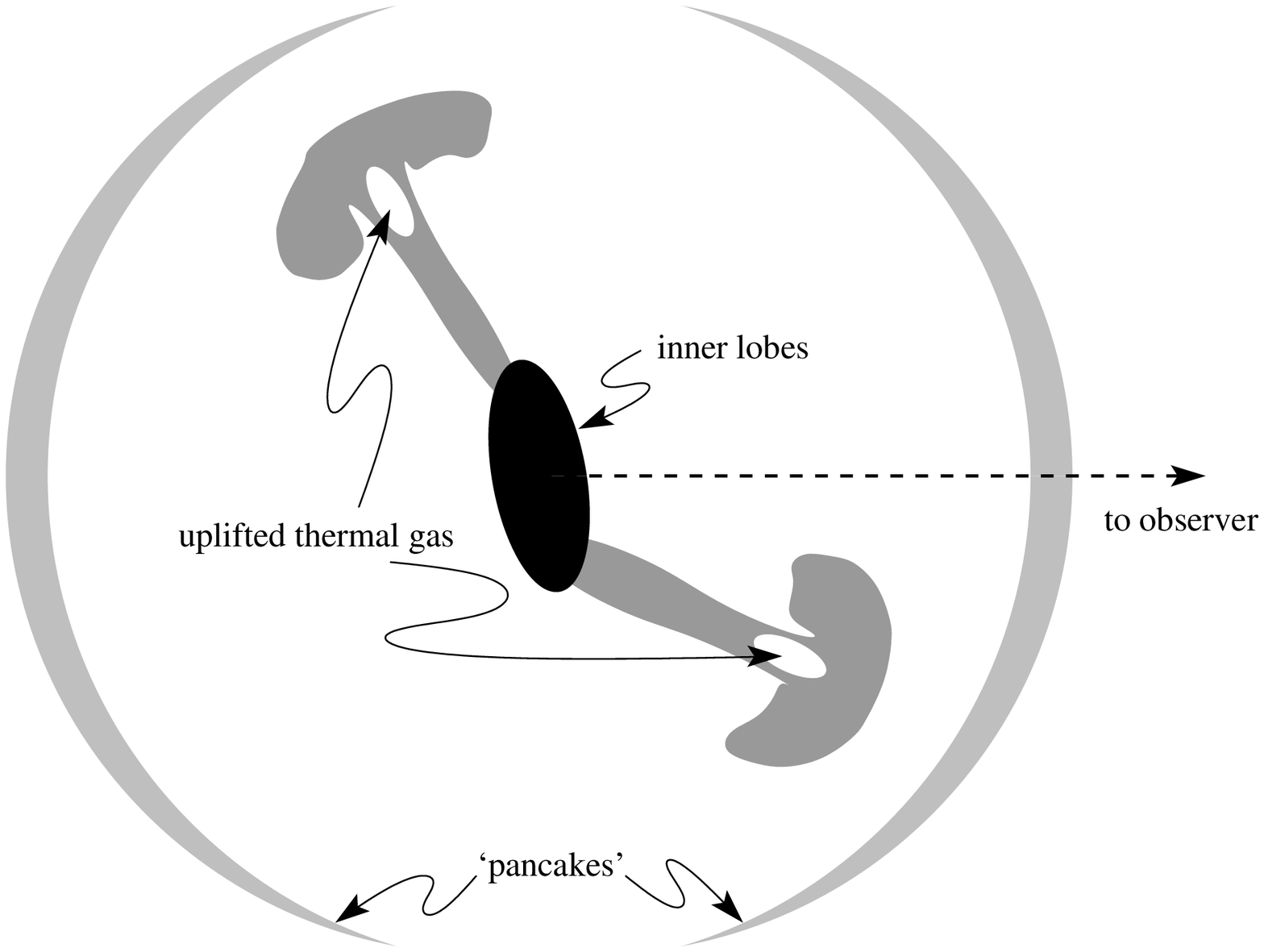}
\caption{
{\bf Top-Left:} $14'.6 \times 16'.0$ map of the radio halo of M87 at
327 MHz rotated  
90\deg clockwise (North is to the right, East is up). The map was kindly
provided by F.Owen (see Owen, Eilek and Kassim 2000 for original data). {\bf 
Top-Right:} Adaptively smoothed ROSAT HRI X--ray image. The size and
orientation of the image are the same as for the radio image.
{\bf Bottom:} Possible geometry of the source inspired by 
analogy with ``mushroom clouds'' produced by powerful atmospheric
explosions. The 
black region in the center denotes the inner radio lobes, gray ``mushrooms''
correspond to the buoyant bubbles already transformed into tori,
and the gray lens-shaped structures are the `pancakes' formed by the
older bubbles. To explain the observed morphology, the source must be
oriented close to the line of sight (dashed line). The pancakes are
shown edge-on as shaded regions.
\label{radio}
\label{fig:m87}
\label{xray}
\label{nuc}
\label{fig:geo}
}
\end{figure*}

\subsection{Radio and X--ray morphology}
\label{sec:obs}

There is a long history of M87 radio halo observations, starting with
Mills (1952) shortly after the initial discovery of the radio source
Virgo A (Bolton, Stanley and Slee 1949). Because of the very bright
compact source, the dynamic range of the images was always a
problem. Recently, Owen et al. (2000) presented a new high resolution,
high dynamic range map of the halo of M87 at 327 MHz (see Figure
\ref{radio}). For the sake of further discussion this map has been
rotated by 90 degrees to have East up and North to the right. In this
map several distinct structures are discernible. The high brightness
centre represents the well-studied inner lobe structure (oriented
approximately east-west) with the famous jet pointing west north-west
(approximately bottom--right for the orientation of images adopted in
Fig.\ref{radio}). The highly structured outer halo is much
fainter. It consists of the ear-shaped (torus-like) eastern bubble,
the much less well-defined western bubble, both of which are connected
to the central emission by a `trunk', and the two very faint, almost
circular, emission regions northeast and southwest of the centre.

Owen et al. (2000) find a peak surface brightness in the region of the
eastern bubble of about 53 mJy/beam while the circular regions range
from 14 mJy/beam to about 25 mJy/beam. Rottmann et al. (1996) show
that all of the described structures are also present at 10.6
GHz. They also show that the outer halos have a radio spectral index
in the range from $-2.8$ to $-2.5$ between 4.8 GHz and 10.6 GHz. The
spectral index is flatter by a factor of about 1.5 between 330 MHz and
1.5 GHz. This suggests that radiative energy losses of the synchrotron
emitting particles lead to a significant steepening of the radio
spectrum at about 3 GHz. Unfortunately, the maps
used by Rottmann et al. (1996) have low resolution since they are
obtained with single dish observations and so the spectral index
distribution of the halo and its connection to the detailed
structures, e.g., bubbles and attached ``trunks'',
is not known.\\

In Fig.\ref{xray} (lower left) we show the X--ray surface brightness
distribution of the same region (and similar orientation) as the
radio map. The data are the combined 200 ksec ROSAT/HRI image,
adaptively smoothed to suppress high frequency noise. As pointed out
by a number of authors (Feigelson et al. 1987, B{\"o}hringer at
al. 1995, Owen, Eilek and Kassim 1999, Harris et al. 1999) there is
evidence for a correlation between X--ray and radio emitting
features. The simplest explanation of this correlation is that the
excess X--ray emission is due to inverse Compton scattering of the
cosmic microwave background photons by the same relativistic electrons
which produce the synchrotron radio emission (Feigelson et
al. 1987). However, ROSAT/PSPC observations have shown that the excess
emission has a thermal spectrum (B{\"o}hringer at al. 1995) and the
X--ray emitting gas in these region has a lower temperature than that
in the ambient regions.

Thus, we summarize three observational facts which we are trying to
explain below:
\begin{itemize}
\item There are prominent ``torus--like'' features in the radio image.
\item There is a correlation (but not one to one) of the X--ray and
radio bright regions.
\item Excess X--ray emission, associated with the radio bubbles and
attached ``trunks'', arises from thermal gas whose temperature is less
than that of the ambient X-ray emitting gas.
\end{itemize}

\subsection{Analogy with atmospheric explosions}
Concentrating on the ``torus--like'' radio features one can note its
striking similarity with some evolutionary stages of hot buoyant
bubbles formed by a powerful (e.g. nuclear) explosion in the Earth's
atmosphere. The transformation of the initially
spherical bubble into a 
torus is a common and well known property of buoyant bubbles
lacking strong surface tension (e.g. Walters and Davison 1963,
Onufriev 1967, Zhidov et al. 1977). 
Morphologically similar structures resembling "mushrooms" appear in
Rayleigh-Taylor unstable configurations: as the fluid rises through the
ambient medium, Kelvin-Helmholtz instabilities create the torus-like
head of the "mushroom".  Images of various laboratory
experiments where such structures were observed and discussion of the relevant
physical mechanisms are widespread (see e.g. Batchelor
1967, Turner 1973, Inogamov 1999). 
The idea that buoyancy plays an important role in the
evolution of the radio lobes in galaxies was first proposed by
Gull and Northover (1973) and has been used to estimate the life time
of the radio lobes in M87 by B\"{o}hringer et al. (1995). The role of
buoyancy was also discussed by Baum and O'Dea (1991) in application to
PKS 0745-191.  
The similarity
of the M87 ``torus--like'' features to the Rayleigh--Taylor mushrooms
was pointed out by Churazov et al. (2000) and to a subsonic vortex
ring by Owen, Eilek and Kassim (2000).

Further pursuing the analogy with powerful explosions, we note that
during the transformation of a bubble into a torus some 
ambient gas is captured and uplifted by the rising
bubble/torus. 
In the cooling flow surrounding M87, the entropy
of the X--ray emitting gas rises quickly with distance from the
centre, and therefore the emissivity of the gas increases very steeply
if it remains in pressure balance with the ambient gas
(B\"{o}hringer et al. 1995, Nulsen 1997). This
may qualitatively explain the correlation of the radio and X--ray
emitting plasmas and naturally accounts for the thermal nature of the
excess thermal emission. \\

In the case of an atmospheric explosion, the final stage of the hot
bubble occurs when the bubble reaches a height at which the density
of the ambient gas is equal to the density of the bubble. Then
(neglecting oscillations near the equilibrium position) the bubble
expands laterally (forming a ``pancake'') to occupy a
thin layer in the atmosphere having the same
density.  The bubbles/tori in M87 may share the same fate. In a
spherically symmetric potential, the bubble will try to fill a segment
(equipotential surface) of a sphere. The largest distinct features
in the radio map (see Fig.\ref{radio}) could be just those late-stage
bubbles.

The sketch of the possible overall source structure based on the
analogy with a powerful atmospheric explosion is shown in
Fig.\ref{fig:geo}. Here the black region in the center denotes the
inner radio lobes, the gray mushrooms correspond to the buoyant
bubbles already transformed into tori and the gray lens-shaped
structures are the ``pancakes'' formed by the older bubbles.

Buoyant bubbles and their role in the complex X--ray and radio
morphology of M87 have been discussed previously e.g. in B\"{o}hringer et
al. (1995), Nulsen (1997), B\"{o}hringer (1999),  Owen, Eilek and Kassim
(2000). Below we follow the line of arguments suggested in Churazov et
al. (2000) and present hydrodynamic calculations of the radio lobe
evolution to illustrate the qualitative picture described
above.

\section{Method}
\label{sec:method}

In the next sections we describe the method, initial conditions, and
assumptions used in the numerical simulations of bubbles in a hot
gaseous atmosphere. We first detail the method and the use of
``tracer'' particles to track the motion of hot gas which is entrained
in the rising bubble. We next describe our approach to calculation of
the synchrotron radio emission from the evolving bubble.

\subsection{Hydrodynamic simulations}

\begin{figure}[t]
\plotone{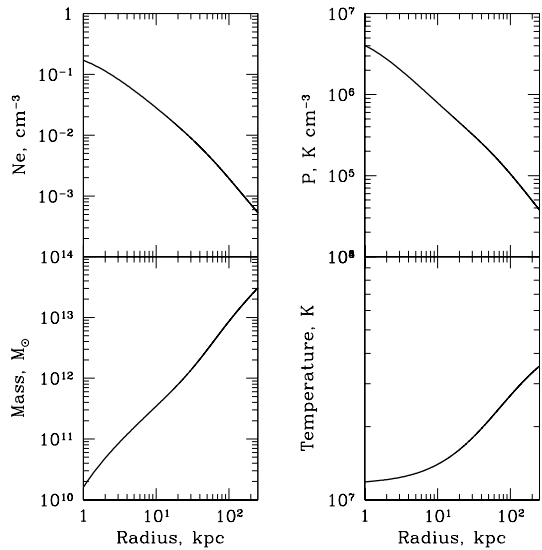}
\caption{Initial distribution of electron density, temperature,
pressure and gravitating mass (adopted from Nulsen and B\"{o}hringer
 1995)
assumed in the simulations.
\label{setup}
}
\end{figure}

The simulations were obtained using the ZEUS-3D code which was
developed especially for problems in astrophysical hydrodynamics
(Stone \& Norman 1992a).  The code uses finite differencing on an
Eulerian grid and is fully explicit in time. It is based on an
operator-split scheme with piecewise linear functions for the
fundamental variables. The fluid is advected through a mesh using the
upwind, monotonic interpolation scheme of van Leer (1977). For a detailed
description of the algorithms and their numerical implementation see
Stone \& Norman (1992a, b). To study the spectral aging of
the relativistic gas, the ZEUS code was modified to follow the motion
of `tracer' particles which are advected with the fluid.\\

In our simulations we employed an ideal gas equation of state and we
ignored the effects of magnetic fields and rotation.  The cooling time
is sufficiently long compared to the time scales considered here
since, for an electron density of 0.02 cm$^{-3}$ and temperature of
the order of 2 keV (see Fig.\ref{setup}) the cooling time is $\sim
5\times 10^8$ years. This is an order of magnitude longer than the
typical duration of the run of $\sim 5\times10^7$ years. We therefore
completely neglected cooling.  The simulations were computed on a
spherical grid in two dimensions and were performed on an IBM RS/6000
cluster. The computational domain spans 51 kpc in radius and $\pi /2$
rad in angle and was covered by 200 $\times$ 200 grid points (in the
$r$- and $\theta$-direction). \\

\begin{figure*}[t]
\leavevmode 
\includegraphics[width=16cm]{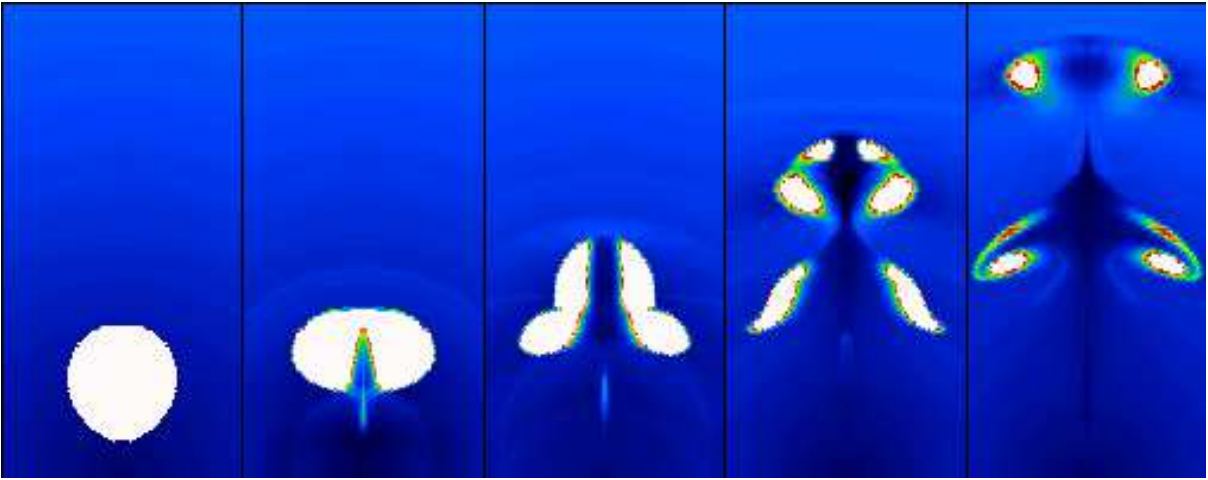} 
\caption{Temperature distribution in the gas at 5 time steps. From
left to right: 0, 8.4, 21, 42 and 67 Myrs after the start of the
simulation. Each box is 40 by 20 kpc. The center of the
cluster is at the bottom of the box. 
Temperature is color coded in the
0.7 (black) to 5 keV (white) range. The temperature of the cluster
thermal gas (blue) changes from $\sim 1$ keV at the center to $\sim 1.7$
keV at a distance of 40 kpc. All temperatures above 5 keV are
white. Thus, the
hot ``radio-emitting plasma'' which  initially has a temperature of
order 100 keV is white.
Numerical diffusion (due to very strong gradients in
density and temperature) creates regions with intermediate values of
temperature (yellow, red, green) along the boundaries of the bubble. 
Note that during later stages, the coldest gas is not at the centre of
the cooling flow, but is associated with the rising bubble. Since
radiative cooling was not included in the simulations these cold
regions represent uplifted (and adiabatically expanded) gas.
\label{snap}
}
\end{figure*}

A model for the mass and initial temperature was adopted from Nulsen
\& B\"ohringer (1995) which is shown in Fig.\ref{setup}. The gas
density distribution was then found by assuming hydrostatic
equilibrium to maintain an initially static model.  A spherical bubble
was created with a radius of $r_{\rm b}=5$ kpc at a distance of $d=9$
kpc from the gravitational center.  It was made buoyant by reducing
its density with respect to the background  by a factor of 100
and simultaneously raising the temperature by the same factor. During
its subsequent evolution, the gas is treated as a single fluid which is
assumed to obey a polytropic equation of state with $\Gamma
=5/3$. The bubble was filled with 2000 tracer particles and was placed
on the $\theta =0$  
axis along which it rose in the gravitational potential. Subsequently,
we rotated the two-dimensional output about this axis
(azimuthally). To test the assumption of rotational symmetry, we
performed some 3D simulations with reduced resolution and found that
this assumption is fairly robust as far as the global motions and
morphologies are concerned. As we discuss in detail below, buoyancy 
deforms the bubble and drives
it through the ambient medium as shown in Fig.\ref{snap}. Shear
instabilities cause the formation of tori which separate from the main
bubble. This process, also called ``vortex shedding'', has been
observed and studied extensively (see Norman et al. 1982). \\

Finally, we should address some issues related to the accuracy of
these kinds of finite-difference hydrodynamic simulations. While the code
can simulate large-scale mixing due to Rayleigh-Taylor and
Kelvin-Helmholtz instabilities, it does not include real particle diffusion. 
Any observed diffusion is therefore entirely numerical. The
boundary between the bubble and the ambient medium becomes less sharp as 
the simulation proceeds due to discretization errors in the
advection scheme. For a test of the advection algorithm in the ZEUS 
code see Stone \& Norman (1992a,b).  In simple advection tests, it was
found that, during the advection of a sharp discontinuity over a grid
of 200 zones, the discontinuity is spread over 3-4 grid
cells. Therefore, the small features in our simulation are likely to
be affected by these advection errors whereas the larger features are
not.

Second, numerical viscosity is also responsible for suppressing
small-scale instabilities at the interface between the bubble and the
cooler, surrounding, X-ray emitting gas.  To assess the effects of
numerical viscosity, we have repeated our simulations on grids with
100 $\times$ 100 grid points. From our experiments, we can conclude
that ``global parameters'' such as the position and size of the bubble
as well as the presence of ``toroidal'' structure are relatively
insensitive to the resolution. The detailed small scale
morphology does depend on the resolution and the initial
conditions.

Finally, we stress that the simulations presented here are not intended
to provide a detailed simulation of the radio/X-ray structure
surrounding M87.  Rather they are intended as a guide, a toy model, to assist
our interpretation of the complex emission (see discussion in Section
3.1).

\subsection{Simulation of the radio emission}
\label{sec:radio}

The observations of radio synchrotron emission from the buoyant
bubbles in M87 imply the presence of a relativistic, magnetized
plasma. However, the simulations presented in previous sections are
based on a purely hydrodynamic scheme with a single,
non-relativistic fluid. To estimate the radio flux expected from the
simulation, we make the following assumptions.

The relativistic plasma and the associated magnetic field, the
`relativistic fluid', are confined within small volumina intermixed
with the non-relativistic, thermal plasma governing the dynamics. This
implies that the relativistic fluid may mix with the thermal fluid on
macroscopic scales but not on microscopic ones. This is very similar
to the behavior of air bubbles in water which are confined by surface
tension (e.g. Zhidov et al. 1977). 

We assume the magnetic field to be tangled on scales small compared to
the size of the volumina of relativistic fluid. This allows us to
treat the magnetic field as part of the relativistic fluid. The total
energy density inside the small volumina is therefore the sum of the
energy density of the magnetic field, $u_{\rm B}$, and that of the
relativistic particles, $u_{\rm e}$. We assume that the small volumina
of relativistic fluid are in pressure equilibrium with the thermal
fluid throughout the simulation, i.e. $3p_{\rm th} = u_{\rm B} +u_{\rm
e}$. We also assume that there is no significant re-acceleration of
relativistic particles in the bubble during its evolution.

At the start of the simulation, we assume some initial value of
the magnetic field (the same value for every small volumina of
relativistic particles within the bubble) and a power law distribution
for the relativistic electrons (in terms of the particle Lorentz
factor), $\gamma$, $n_{\rm e} ( t_{\rm o}) d\gamma = n_{\rm o} \gamma
^{-p} d\gamma$. At the start of the simulation $u_{\rm e}$ is
therefore given by 
\begin{equation}
u_{\rm e} = m_{\rm e} c^2 n_{\rm o} \int_{\gamma _{\rm min}}^{\gamma
_{\rm max}} \gamma ^{-p} \left( \gamma -1 \right) \, d\gamma,
\label{power}
\end{equation}
\noindent where we assume $\gamma _{\rm min}=1$, $\gamma _{\rm
max}=10^6$. For a given initial 
magnetic field 
the normalization $n_{\rm o}$ of the particle distribution is then
adjusted such that the entire relativistic fluid is in pressure
equilibrium with the thermal fluid.

For values of $p\ge 2$, the energy density of relativistic particles
is dominated by low energy particles. The life time of such particles,
due to radiative losses, is much longer than the duration of the
simulation (and no Coulomb losses are present since we assume that
thermal gas is not mixed with relativistic particles on microscopic
scales). The motion of the bubble (see Section 3) is found to be
subsonic even with respect to the cluster gas. We therefore assume
that the energy density of the magnetic filed and relativistic
particles changes adiabatically according to the changes of the
pressure of the surrounding thermal gas: $u_{\rm B} \propto u_{\rm e}
\propto p_{\rm th}$. Note, however, that our hydrodynamic simulations
assume an adiabatic index of 5/3 for the nonrelativistic gas, while
the relativistic fluid expands and contracts adiabatically with an
adiabatic index $\Gamma _{\rm r} =4/3$. This implies that, during the
expansion of a buoyant bubble, the fraction of the volume occupied by
the relativistic fluid increases relative to the volume of the thermal
fluid as $V_{\rm r} \propto V_{\rm th}^{5/4}$. Strictly speaking this
means that accounting for relativistic fluid with a different
adiabatic index will not affect the dynamics of the bubble evolution
as long as the fraction of the volume occupied by the relativistic
fluid is small. We will see in the following that this assumption is
probably invalid, but we believe that it does not very severely affect
the results.

The energy losses of the relativistic particles are due to adiabatic
expansion, synchrotron radiation and inverse Compton scattering of the
CMB. These losses depend on the pressure history of each volumina
of relativistic particles. The pressure history can be
recovered using the tracer particles. For a given region within the
bubble the change of the Lorentz factor $\gamma$ of a given electron
due to adiabatic expansion or contraction of the relativistic fluid
and radiative losses can be written as  

\begin{eqnarray}
\frac{d\gamma}{dt} & = & a \frac{\gamma}{p_{\rm th} (t)} \,
\frac{dp_{\rm th} (t)}{dt}\nonumber \\
& - & \frac{4 \sigma _{\rm T}}{3 m_{\rm e}c} \gamma ^2 \left[ u_{\rm
B} (t_{\rm o}) \frac{p_{\rm th} (t)}{p_{\rm th}(t_{\rm o})} + u_{\rm
IC} \right],
\end{eqnarray}
where $a=(\Gamma-1)/\Gamma=1/4$. 

\noindent This can be integrated  (e.g. Kaiser, Dennett-Thorpe \&
Alexander 1997) to give the relation between the Lorentz factor of the
electron at time $t$ and the initial Lorentz factor of the same electron

\begin{equation}
\gamma_0 = \frac{\gamma~p_{\rm th}
(t)^{a}}{p_{\rm th} (t_0)^{a} + \gamma~b},
\label{iter}
\end{equation}

\noindent with

\begin{eqnarray}
b=\frac{4 \sigma _{\rm T}}{3 m_{\rm e}c} \int_{t_{\rm o}}^{t} \left[ u_{\rm
B} (t_{\rm o}) \frac{p_{\rm th} (t)}{p_{\rm th}(t_{\rm o})} + u_{\rm
IC} \right] p_{\rm th}
(t)^{a} dt.
\end{eqnarray}

\noindent The value of $b$ is calculated  from the pressure history of
the tracer particles. The integral in this expression
must be evaluated numerically using the discrete time steps of the
simulation. 

Finally, the adiabatic change of the volume in which the
relativistic particles 
are contained  implies $n_{\rm o} (t) = n_{\rm o} (t_{\rm o})
[p_{\rm th} (t) / p_{\rm th} (t_{\rm o}) ]^{1/\Gamma}$. The energy
distribution of relativistic electrons at time $t$ is then

\begin{equation}
n_{\rm e} d\gamma = n_{\rm o} (t)  \gamma_0 ^{-p} \frac{d\gamma_0}{d\gamma}d\gamma=
n_{\rm o} (t_{\rm o}) \frac{p_{\rm th}
(t)\gamma_{\rm o}^{2-p}}{p_{\rm th} (t_{\rm
o})\gamma^2} d\gamma.
\end{equation}

The energy distribution of the relativistic particles combined with
the strength of the magnetic field yields the monochromatic
synchrotron emissivity (e.g. Longair 1994)

\begin{equation}
j_{\nu}=\frac{4}{3} \sigma_{\rm T} c u_{\rm B} \int _{\gamma _{\rm min}} ^{\gamma _{\rm max}} \gamma ^2 \phi (\gamma, \nu)\, n_{\rm e}\, d\gamma,
\end{equation}

\noindent where $\phi (\gamma, \nu)$ is the dimensionless spectrum
emitted by a single electron.

To obtain the synchrotron surface brightness from the simulation, we
assume rotational symmetry about the polar-axis and integrate the
emissivity along lines of sight through the resulting 3-dimensional
model.

\begin{figure*}[ht]
\leavevmode 
\includegraphics[width=16cm]{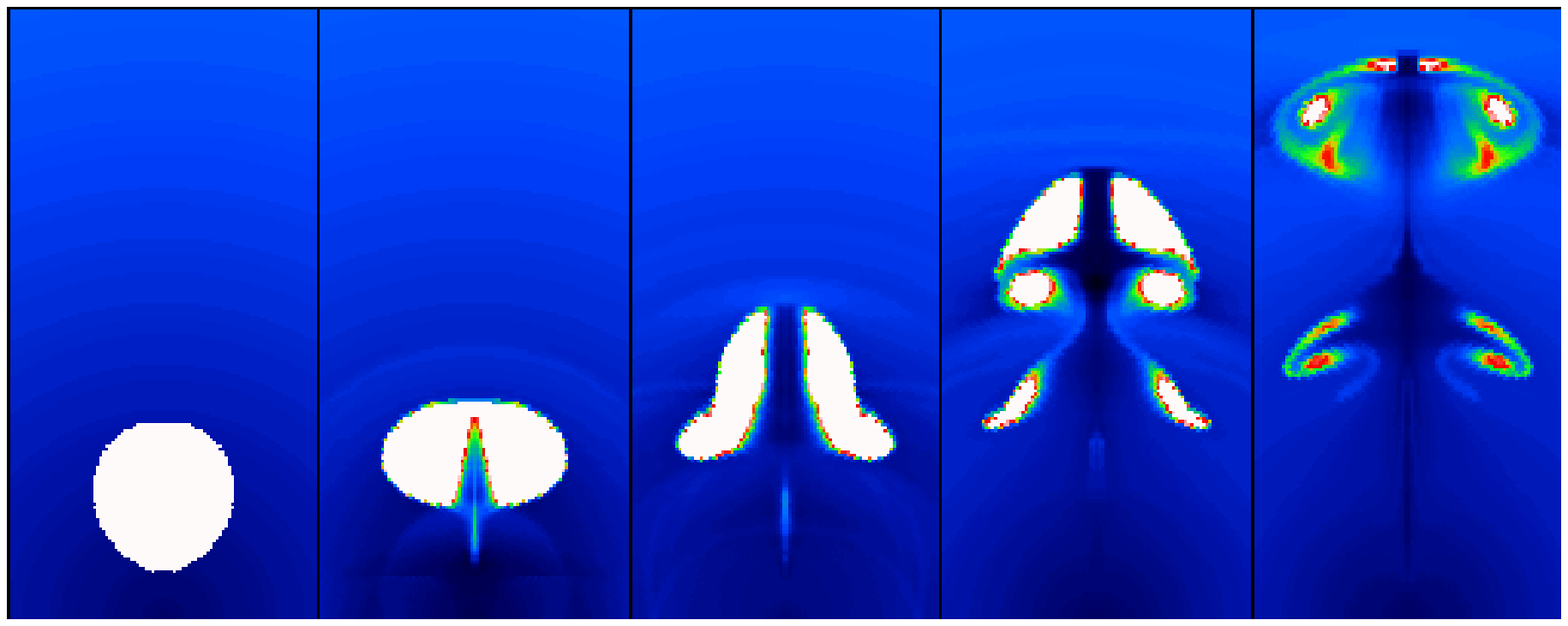} 
\caption{The same as in the previous figure but with an initial
temperature contrast, between the ambient gas and the bubble, which is
a factor of two larger. Note that detailed morphology of the rising
bubble is rather sensitive to the initial conditions (and boundary
conditions, symmetry assumed etc. ) and can be very different from
what is shown in these figures. The presence of the torus--like
structure is, however, a generic feature of the rising bubble
phenomenon. 
\label{snap2}
}
\end{figure*}

\section{Results and discussion}

With the foundation and assumptions described in the previous
section, we describe the results of two numerical simulations. We
first present the results from the two numerical simulations with
different initial conditions for the bubbles. We compute the velocity
and position versus time of the rising bubble. For the first
simulation, we compute the synchrotron radio emission from the bubble
for two orientations, one aligned with the symmetry axis in the plane
of the sky and the second withthe symmetry axis aligned 45$^{\circ}$
with respect to the plane of the sky.  The two orientations differ in
the initial conditions for the magnetic field strength in order to
reproduce the observed radio spectrum. We derive the X-ray appearance
of the bubble from the first simulation, also for the two different
orientations. Finally, we discuss the ``pancake'' phase and calculate
the largest distance to which bubbles can rise as a function of their
initial location and volume fraction of trapped, hot gas.

\subsection{Bubble configuration}

Fig.\ref{snap} shows  snapshots of the temperature
distribution during the evolution of the bubble. One can identify two
stages of the bubble evolution. The initially spherical bubble flattens
and develops a ``cone'' at the rear, which is filled with entrained
 gas. The bubble then transforms into a torus which at later
stages may fragment into smaller structures, but which continue to
show a global toroidal structure. Note that ambient gas (in
particular the gas captured during the bubble $\rightarrow$ torus
transformation) occupies the central part of the rising
structure. Note also that in the last stages (shown in Fig.\ref{snap})
the coldest temperatures are found not at the center of the cluster
but within the rising structure. We stress again that no radiative
cooling was included in the simulations and these low temperatures
regions are associated with gas which has been uplifted from the
central regions and has expanded adiabatically to match the ambient
gas pressure at the current location of the bubble.\\

Of course the particular shape of the torus/mushroom structure
(e.g. two tori at different distances from the center in
Fig.\ref{snap}) is the result of our choice of initial and boundary
conditions and assumed axial symmetry. For example, Fig.\ref{snap2} shows
snapshots of the temperature distribution for slightly
different initial conditions (with an initial temperature contrast,
between the ambient gas and the bubble, which is a factor of two
larger than in the previous run). Note the changes in the detailed
morphology of the rising bubble compared to Fig.\ref{snap}.
In general one would expect a large variety of structures to be formed,
especially in 3D. But the torus-like geometry is a generic feature of
rising bubbles in the absence of strong surface tension.

\subsection{The velocity of the rising bubble}
\label{sec:veloc}
\begin{figure}[t]
\plotone{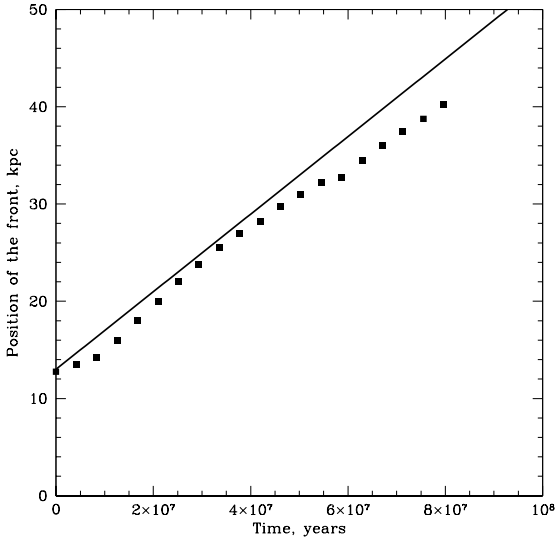} 
\caption{Position of the bubble front as a function of time. For
comparison solid line shows the motion with the constant velocity of
390 km/s.
\label{vrise}
}
\end{figure}

The simplest velocity estimate of the rising bubble can be
obtained by equating the ram pressure and buoyancy forces acting upon
a bubble (e.g. Gull and Northover 1973). The
buoyancy force is obviously:
\begin{eqnarray}
F_{\rm b}=Vg(\rho_{\rm a}-\rho_{\rm b}),
\end{eqnarray}
where $V$ is the bubble volume, $g$ is the gravitational acceleration
(we assume the ambient gas is in hydrostatic equilibrium),
$\rho_{\rm a}$ and $\rho_{\rm b}$ are the mass densities of the
ambient and the bubble gas respectively.  The ram pressure
(drag) is:
\begin{eqnarray}
F_d\sim C \frac{1}{2} Sv^2\rho_{\rm a},
\end{eqnarray}
where $S$ is the cross section of the bubble. 
The numerical coefficient $C$ (drag coefficient) depends
on the geometry of the bubble and the Reynolds number. 
Thus the terminal velocity of the bubble is
\begin{eqnarray}
v\sim\sqrt{g\frac{V}{S}\frac{2}{C}\frac{\rho_{\rm a}-\rho_{\rm b}}{\rho_{\rm a}}}\sim\sqrt{g\frac{V}{S}\frac{2}{C}}.
\label{eqvr}
\end{eqnarray}
Here the factor of $(\rho_{\rm a}-\rho_{\rm b})/\rho_{\rm a}$ 
can be dropped if the bubble density is low compared to the ambient
gas density. The expression for the terminal velocity can be further
rewritten using the Keplerian velocity at a given distance from the
cluster center: $v\sim \sqrt{(r/R)(8/3C)}v_{\rm K}$, where
$r$ is the bubble radius, 
$R$ is the distance from the center and $v_{\rm K}=\sqrt{gR}$ is the
Keplerian velocity. For a solid sphere moving through an
incompressible fluid the drag coefficient $C$ is of the order 0.4--0.5, for
Reynolds numbers in the range $\sim 10^3$--$10^5$ (e.g. Landau and
Lifshitz 1963). The drag coefficient for a rising bubble should of
course be different. First of all, it is not a solid sphere, but a
rather complicated structure, resembling a smoke ring in air. Such rings may
travel large distances through the air with low drag. Secondly the
Mach number of the rising bubble in our simulations is $\sim$0.6--0.7 and
the compressibility of the gas is important. This would tend to increase
the drag coefficient, e.g., for a sphere moving at $M\sim0.7$ the drag
coefficient is $C\sim 0.6$. Thirdly, in a stratified medium an additional
contribution to the drag may come from the excitation of internal gravity
waves. In our simulations, the typical Keplerian velocity was $\sim$ 400
km s$^{-1}$. Fig.\ref{vrise} shows  the position of the bubble front as a
function of time. The solid line shows the expected position of the
feature moving with the constant velocity of $\sim $ 390 km s$^{-1}$. 
The effective
drag coefficient which can be estimated from the simulations,
assuming the above formula for the rise velocity, is $C\sim 0.75$. We
note here that this value may in turn be affected by the numerical
resolution adopted here. However, since the velocity of the bubble depends
on the square root of the drag coefficient,  formula
(\ref{eqvr}) still yields crude order of magnitude
estimates of the bubble velocity. Thus, a large and strongly underdense
bubble will rise with a velocity comparable to the Keplerian velocity.

 In the case of M87,  the bubble's environment may not be very
underdense compared to the ambient gas (since no clear X--ray holes
have been seen) and the velocity of the bubble may be somewhat
smaller. However, we believe that the 
above estimate of the velocity is approximately correct since the
velocity depends only weakly (as the 0.5 power) on the bubble size and
gas density gradients. This velocity defines a time scale of $\sim$
few $10^7$ years for the evolution of the bubble to the stage seen as
a bright torus in Fig.\ref{radio}. 

\subsection{The rising bubble in synchrotron emission}
\label{sec:sync}

\begin{figure*}[t]
\leavevmode 
\includegraphics[width=16cm]{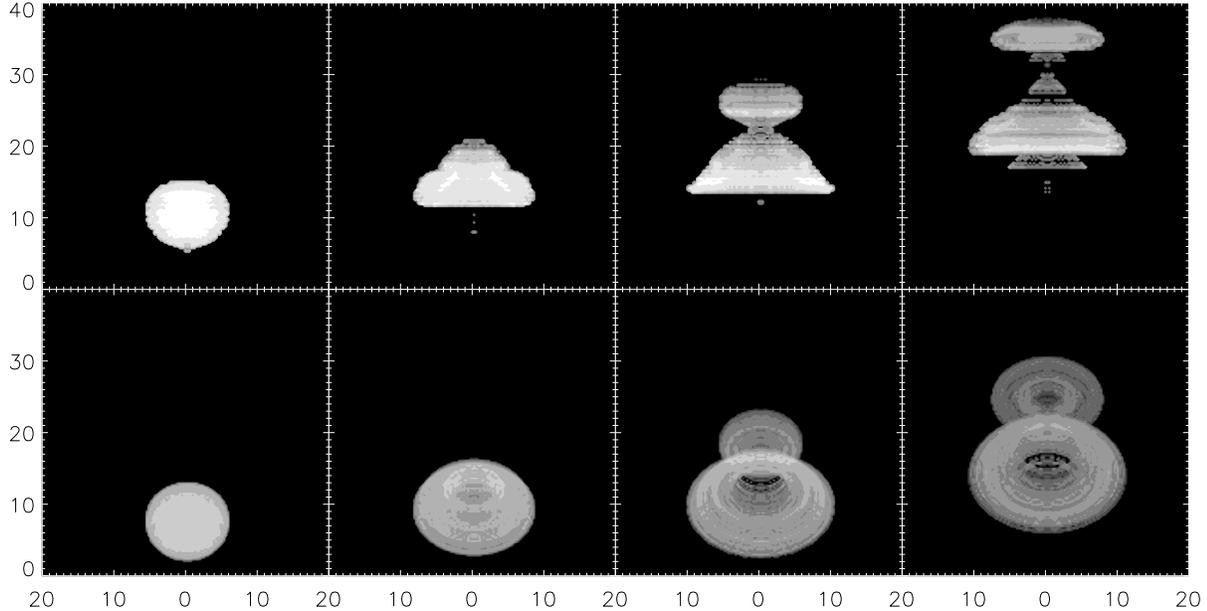} 
\caption{Synchrotron surface brightness at 327 MHz at four time steps
for the simulation shown in Fig.\ref{snap}. The upper row (case 1)
shows the buoyant bubble with a symmetry axis in the
plane of the sky and $ B \sim 6.5 \mu$G initially. The lower row (case
2) is for a symmetry axis inclined 45$^{\circ}$  to the
line of sight and $ B \sim 2.9 \mu$G initially. From left to right
both rows show the bubble at 4.2, 21, 42 and 67 Myrs after the start
of the simulation. The filled contours bound regions of 0.017 (black),
0.04, 0.09, 0.2, 0.45, 1.0, 2.4, 5.5, 13, 28 and 66 (white) mJy per
pixel, where a pixel is 250 $\times$ 250 pc at the distance of M87
($\sim$18 Mpc).}
\label{fig:sync}
\end{figure*}

In Figure \ref{fig:sync} we show the simulation results at 327
MHz. The radio emission was calculated  for the hydrodynamic
simulations shown in Fig.\ref{snap}. Upper and lower rows correspond
to different viewing angles. The symmetry axis is perpendicular to
the line of sight in the upper row (case 1) in this figure and
at an angle of $45^{\circ}$ to the line of sight (case 2) in the
lower row.  From the
radio observations of M87 we note that the well-defined eastern bubble
is observed at a distance of roughly 25 kpc from the source
centre. This corresponds to the third panel in the upper row of Figure
\ref{fig:sync}, i.e. bubble age of 42 Myrs, while for the
simulation projected at an angle of 45$^{\circ}$ (lower row), the
fourth panel, at an age of 67 Myrs, gives the correct projected
distance. Note here that these distances are for the primary bubble/vortex
ring. Figure \ref{fig:sync} clearly shows the presence of a secondary
vortex ring which is closer to the centre. The appearance of
this secondary ring is fairly dependent on the particular choice of
initial conditions (see Section
\ref{sec:method} and Figures \ref{snap},\ref{snap2}). In the case
studied here, the synchrotron surface brightness of the secondary
vortex ring exceeds that of the primary bubble. No such structure is
observed in M87 and so we concentrate in the following discussion on
the primary bubble.\\ 

For an initial power law distribution of relativistic electrons in a
constant magnetic field, a significant steepening of the synchrotron
spectrum will be observed for frequencies higher than the break
frequency, $\nu _{\rm break}$, which is given by (e.g. Leahy 1991)
\begin{equation}
\nu _{\rm break} = 2.52 \times 10^{6} \frac{B}{\left(B^2+B_{\rm CMB}^2
\right) ^2 t^2} {\rm \, GHz},
\label{break}
\end{equation}

\noindent where $B$ is the strength of the magnetic field in $\mu$G
and $t$ is the time (in Myrs) for which the plasma was exposed to this
field.  $B_{\rm CMB}$ is the strength of the magnetic field in $\mu$G
equivalent to the energy density of the CMB, $u_{\rm IC}$. For the
redshift of M87, $u_{\rm IC} \sim 4.2 \times 10^{-13}$ erg cm$^{-3}$
and so $B_{\rm CMB} \sim 3.2 \mu$G. From equation \ref{break} the
time $t$ can be expressed as a function of the break frequency and
the strength of the magnetic field. For any given break frequency, the
time $t$ is maximal if $B (t_{\rm max}) = B_{\rm CMB} /\sqrt{3} \sim 2
\mu$G. We have seen that the observed radio spectral indices (Section
\ref{sec:obs}) require $\nu _{\rm break} \sim 3$ GHz, for which a
maximum age is $t_{\rm max}\sim 10^8$ years. This is comparable with
the time span of our simulations. For example, if the symmetry axis is at
angle of $45^{\circ}$ with the line of sight (case 2 above), then the
front of the bubble will reach a projected distance of $\sim$25 kpc after
67 Myrs. This simple estimate shows that we require a
rather low value of the initial magnetic field strength (comparable to
$B (t_{\rm max}) \sim 2 \mu$G) in order to avoid having the $\nu _{\rm
break}$  lie at a frequency lower than $\sim 3$ GHz.

For case 1, we used $p=2.3$ and an initial magnetic field strength of
6.5 $\mu$G. We then calculated the synchrotron emission of the bubble
as described in Section \ref{sec:radio}.  After $\sim$ 42 Myrs,
i.e. when the bubble is $\sim$25 kpc away from the center, the
spectral index of the primary bubble is $-2.7$ from 4.7 -- 10.6 GHz
and $-1.0$ from 330 -- 1500 MHz
which compares well with the observed values  (Rottmann et al. 1996,
Section \ref{sec:obs}).

For case 2 the magnetic field inside the bubble must be lower since
here the bubble is older when its upper edge reaches the correct
projected height from the cluster centre. We set $B=2.9 \mu$G and
$p=2.3$. For this case the high frequency spectral index is found to
be $-3.0$ while at lower frequencies it is $-1.0$.

The assumption of pressure equilibrium determines the initial energy
density of the relativistic particles. The energy
density of the magnetic field is $1.7 \times 10^{-12}$ ergs cm$^{-3}$
for case 1 and $3.4 \times 10^{-13}$ ergs cm$^{-3}$ for case 2. This is
much lower than the energy density of the thermal cluster gas and the
particles energy density is therefore $u_{\rm e}\sim 3p_{\rm th} \sim 4.2
\times 10^{-10}$ ergs cm$^{-3}$ for both cases discussed.
This implies a severe departure from the equipartition condition of $u_{\rm
B} \sim u_{\rm e}$ often assumed for synchrotron emitting plasmas. 
The initial volume filling factor of the radio emitting plasma in the
initial bubble can then be adjusted to provide the required surface
brightness of the final bubble. The peak surface brightness predicted
in the region of the primary vortex ring is 55 mJy/pixel in case 1
(see Owen et al. 2000, Section \ref{sec:obs}), however, the required
volume filling factor of the relativistic plasma is very close to
unity. For case 2, even for a filling factor of unity,
the peak surface brightness of the bubble is only 2.4
mJy/pixel. Furthermore, we neglected any possible contribution from the 
relativistic ions (the so-called "k" factor) and thermal plasma to the
pressure inside the bubble, which would further reduce its synchrotron
emissivity. We also note here that volume filling factors $\sim$1 pose
a serious problem for our assumption that the relativistic fluid does
not influence the overall dynamics of the buoyant bubble.\\ 

The problems discussed above arise from the life time of the
synchrotron emitting particles being short compared to the dynamical
time scale of the structure. Gull \& Northover (1973) first noted this 
lifetime problem  for buoyant models. Below, we
discuss various solutions.

From Figure \ref{fig:sync}, we find that the diameter of our primary
bubble is smaller than the observed diameter of the eastern bubble
in M87. As we showed in Section \ref{sec:veloc}, the rise time
of the bubble is correlated with its size. It is therefore quite
possible that the eastern bubble in M87 is somewhat younger than our
simulations suggest. Such a younger age would allow a
stronger magnetic field in the bubble and could explain the
observed spectrum, which in turn would allow a smaller volume
filling factor for the relativistic fluid. 

The buoyant bubble may also be filled mainly by a weak magnetic field
interspersed with regions of a much higher field strength but small
volume filling factor. The relativistic electrons may then survive for
a much longer time in the mainly weak magnetic field and only diffuse
slowly into the regions of greater field strengths which are
responsible for most of the observed synchrotron emission (Eilek,
Melrose \& Walker 1997). Subsonic turbulence can also enhance the
energy density of the magnetic field in regions with initially
weak field (e.g. Eilek, Owen and Zhou 1999). 

If the magnetic field in the buoyant bubbles is stronger than
that estimated above and rather homogeneous, then {\em in situ} acceleration
of relativistic particles is required. In this case, the age of the
buoyant bubble cannot be constrained from radio observations. The two
latter mechanisms were also discussed by Owen et al. (2000) to explain
the radio emission of the outer M87 halo.

The fundamental conclusion of these simulations is that one can
reproduce the observed radio brightness and the break in the spectrum
of the torus--like feature in M87 assuming that (i) the bubble, filled
with the relativistic plasma, is in pressure equilibrium with the
thermal gas, (ii) the initial magnetic field strength is relatively
low so as to prolong the lifetime of the electrons and (iii) no {\em
in situ} reacceleration of particles or generation of magnetic field
occurs during the bubble evolution. However, the equipartition
condition is  sacrificed and the parameter space which satisfies these
assumptions is rather tight. It is likely that the real situation is
much more complex than was assumed in the simple model discussed
above.

\subsection{The rising bubble in X--ray emission}

\begin{figure}[t]
\plotone{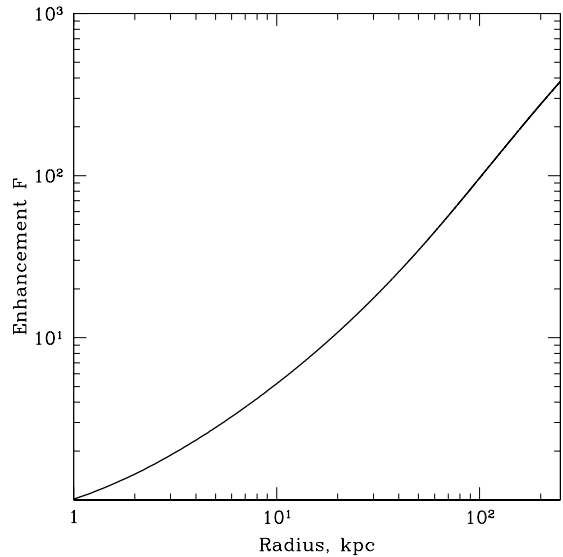}
\caption{Volume emissivity of the uplifted and adiabatically expanded
gas relative to the ambient gas. The curve is normalized to unity at 1
kpc. The enhancement factor, $F(r)$, is the ratio of the emissivity of the
gas uplifted from 1 kpc to $r$ to that of the ambient gas at $r$.
The ratio $F(r_2)/F(r_1)$  characterizes the factor by which
volume emissivity of the gas uplifted from $r_1$ to $r_2$ exceeds the
volume emissivity of the ambient gas at $r_2$.
\label{xenh}
}
\end{figure}

As is clear from Fig.\ref{snap}, the bubble captures ambient gas during
its transformation to a torus and then uplifts this gas to larger
radii. Since all motions in our simulations are subsonic, this uplifted
gas remains in approximate pressure equilibrium with the ambient gas at
larger radii. The volume emissivity of this gas can be compared with
the volume emissivity of the ambient gas:
\begin{eqnarray}
\epsilon_u=n_u^2 \Lambda(T_u) \nonumber \\ 
\epsilon_a=n_a^2 \Lambda(T_a),  
\end{eqnarray}
where $n$ and $T$ are the density and temperature of the uplifted and
ambient gas respectively, $\Lambda(T)$ is the emissivity in a given
energy band. If we assume adiabatic evolution of the entrained and
uplifted gas then:
\begin{eqnarray}
n_u= n_{a,0} \left ( \frac{P_a}{P_{a,0}}\right )^{1/\Gamma}=n_a \frac{T_{a}}{T_{a,0}}\left ( \frac{P_a}{P_{a,0}}\right
)^{1/\Gamma-1}
\end{eqnarray}
\begin{eqnarray}
T_u= T_{a,0} \left ( \frac{P_a}{P_{a,0}}\right )^{1-1/\Gamma},
\end{eqnarray}
where the subscript $0$ stands for the values of the uplifted gas at its
initial location where it was captured by the rising bubble. 
Thus the uplifted gas will have an emissivity which is higher than the ambient
gas by a factor:
\begin{eqnarray}
F=\frac{\epsilon_u}{\epsilon_a}=\left ( \frac{T_{a}}{T_{a,0}} \right )^2 \left ( \frac{P_a}{P_{a,0}}\right
)^{2/\Gamma-2} \times \frac{\Lambda(T_u)}{\Lambda(T_a)}.
\end{eqnarray}
The enhancement factor $F$ (excluding the cooling functions) is shown in
Fig.\ref{xenh}. One can see that in a cooling flow (density 
decreases with distance while temperature increases), the emissivity of the
uplifted gas greatly exceeds the emissivity of the ambient gas if the
displacement is large. The ratio $\Lambda(T_u)/\Lambda(T_a)$ 
(when the temperatures are near 1 keV and the soft X--ray band is
considered) further increases the brightness of the uplifted gas with
respect to the ambient gas. 
\begin{figure*}[t]
\leavevmode 
\includegraphics[width=16cm]{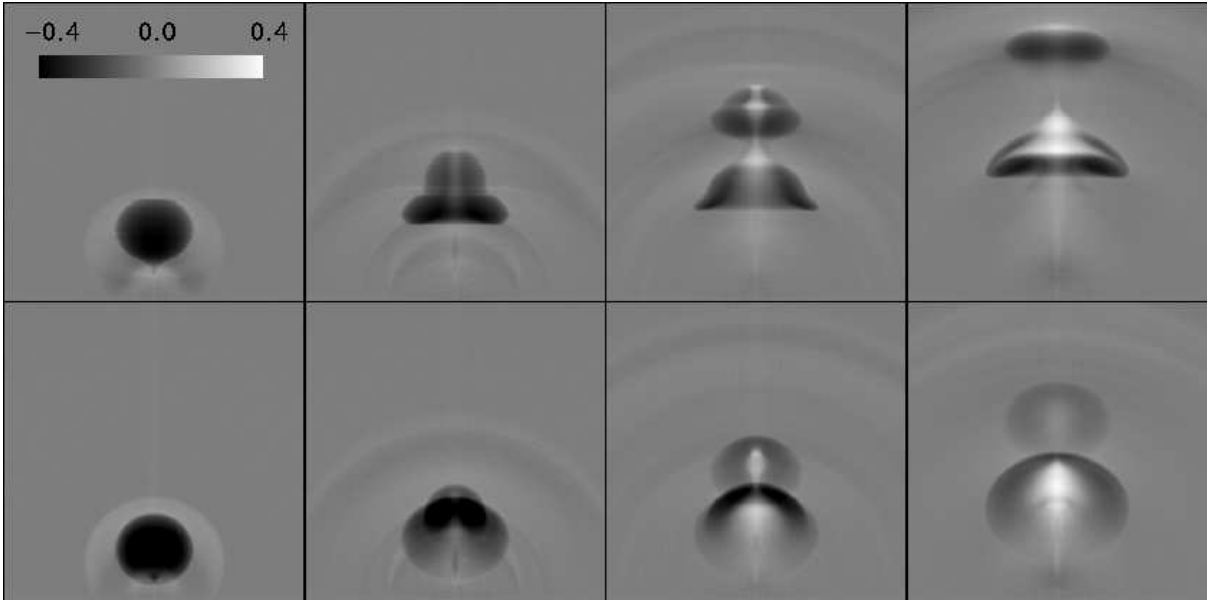} 
\caption{Expected X--ray morphology of the bubble for the same times
as in Fig.6. Each box is 40 by 40 kpc. The figure shows the relative deviation
of the X--ray surface brightness (in the ROSAT energy band) with
respect to the unperturbed X--ray emission of the cooling flow. The
darker regions are X--ray underluminous and lighter regions are X--ray
bright. Black corresponds to regions dimmer by 40\% (or more)
than the unperturbed value and white  corresponds to regions
brighter by 40\% (or more) than the unperturbed value.
\label{xmor}
}
\end{figure*}

\begin{figure*}[ht]
\centering 
\leavevmode 
\includegraphics[width=11cm]{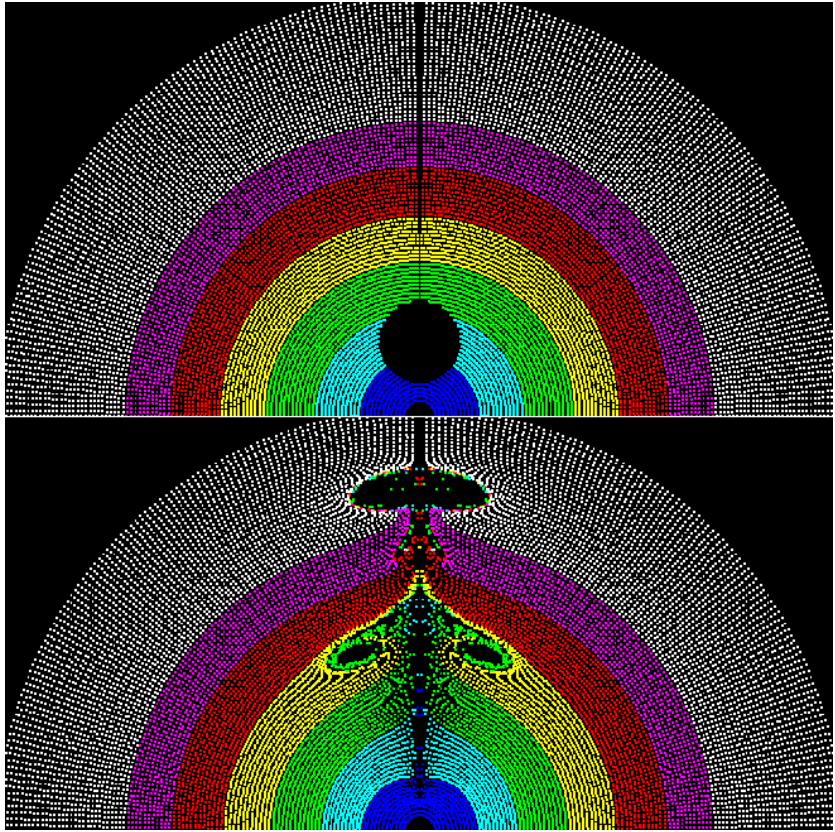} 
\caption{Initial (top) and final (bottom) positions of tracer
particles, originating within different regions of the cluster
gas. Each particle is color coded according to its initial distance
from the center. Bottom panel shows the positions of particles
$\sim$67 Myrs after the beginning of the simulation. ``Green'' and
``blue'' particles seen in the upper torus (bottom panel) have been
uplifted from the central region of the cooling flow.
\label{uplift}
}
\end{figure*}

To illustrate how the X--ray surface brightness is affected by
the a radio emitting plasma bubble and lumps
of uplifted cold gas, we calculated the projected surface brightness
distribution (for the ROSAT energy band from 0.5 to 2 keV) for several
stages of the bubble evolution. When calculating these maps, we assumed
axial symmetry and assumed that the axis of the bubble is either
perpendicular to the line of sight (upper row of images in
Fig.\ref{xmor}) or that the angle between the line of sight and bubble
axis is 45$^{\circ}$ (lower row of images in Fig.\ref{xmor}).
Fig.\ref{xmor} shows the relative deviation of the X--ray surface
brightness from the unperturbed value of the surface brightness at the
same distance from the center. As expected the ``trunk'' of the
mushroom is brighter than the surrounding regions.

\begin{figure}
\plotone{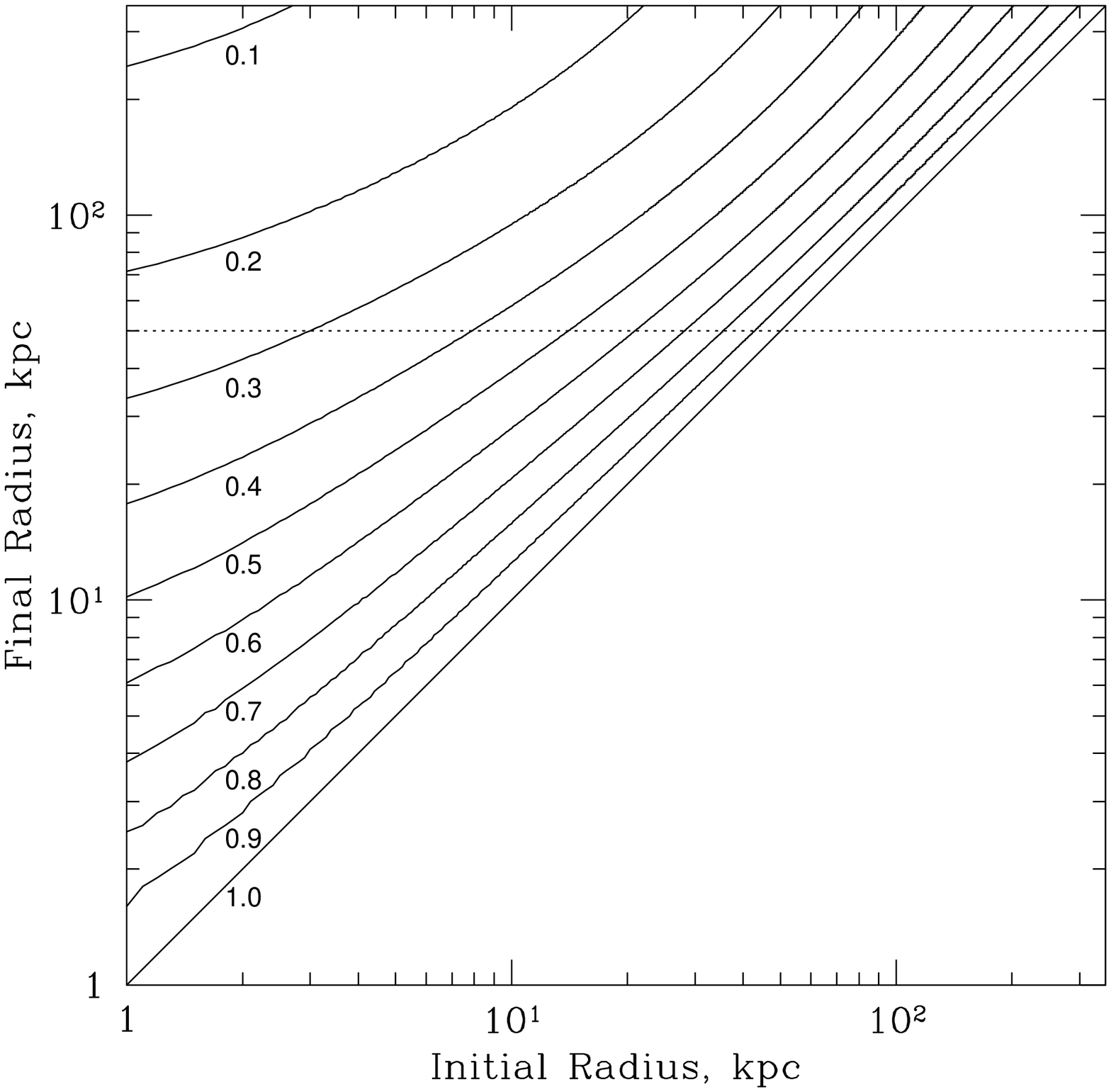}
\caption{The maximum radius to which the bubble can rise depending on its
initial location and the
volume fraction of gas, captured during the transformation of the bubble
into a torus. It is assumed that the bubble 
consists of separate lumps of radio and X--ray emitting
plasmas bound together.  Each curve corresponds to a given volume
fraction (labels under the curves) of entrained gas added to the
bubble near the original bubble position. If, for example,
the large circular radio emission regions are located at a
distance of 50 kpc from the center (horizontal dotted line) then a  
bubble could rise from 5 kpc to 50 kpc with the amount of
``entrained'' gas corresponding to
approximately 35\%  of its initial volume.
\label{rfin}
}
\end{figure}

Fig.\ref{uplift} illustrates the effect of entrainment of cold gas as
the bubble rises. Shown in this figure are the distributions of tracer
particles at the initial and final (after $\sim$67 Myrs) stages of the
bubble evolution. The tracer particles fill the entire volume outside
the bubble. The color assigned to each particle characterizes its
initial distance from the cluster center. For example, ``green'' and
``blue'' particles seen in the upper torus have been uplifted from the
central region of the galaxy or cluster atmosphere.

Abundance gradients are frequently observed in cluster atmospheres
around central bright galaxies (e.g. Matsumoto et al. 1996). If our
interpretation of the X--ray features as uplifted lumps of gas is
correct, then one would expect the abundance in these regions to be
higher than in the surrounding regions. This can be tested with
Chandra and XMM.

\subsection{The pancake stage}

If the lumps of radio and X--ray emitting plasmas in the bubble are
bound together (e.g. by magnetic stresses as in the multiphase
cooling flow picture of Nulsen (1986)) then one can calculate the
maximum radius to which the whole bubble can rise for given volume
fractions of ``entrained'' gas and (weightless) radio emitting plasma
(see eq. (6) in Churazov et al. 2000). Fig.\ref{rfin} shows the
dependence of the final position of the bubble on the initial position
and fraction of the bubble volume occupied by entrained gas
which was captured near the initial location of the bubble. At this final
location, the average mass density of the bubble (consisting of separate lumps
of radio and X--ray emitting plasma) is equal to the density of the
ambient gas. At the final position the bubble will spread along the
equipotential (isodensity) surface and may form a pancake--like
structure. The X--ray appearance of the bubble at this stage
crucially depends on whether mixing of radio and X--ray emitting
plasmas is microscopic or macroscopic (B\"ohringer et al. 1995, Nulsen 1997)

\begin{figure*}[t]
\centering 
\leavevmode 
\includegraphics[width=13cm]{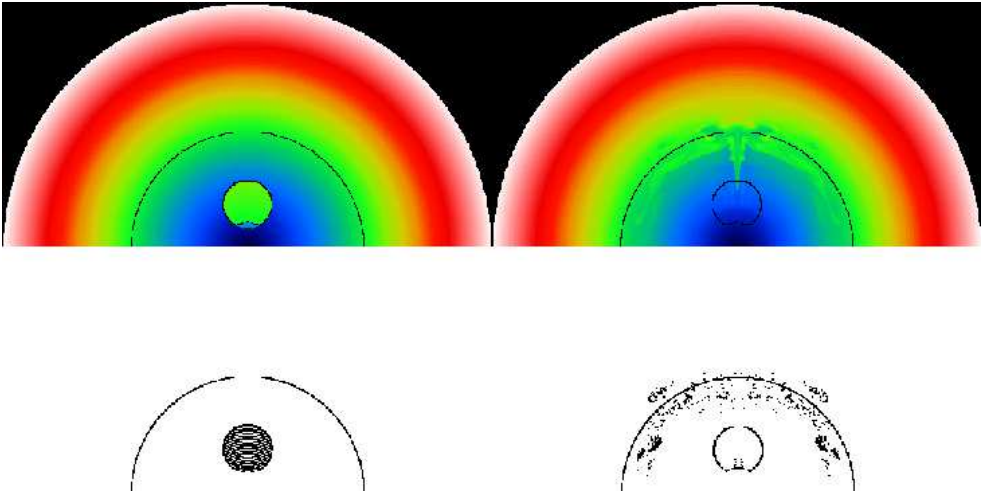} 
\caption{Simple simulations demonstrating formation of a pancake. The
top panel shows the distribution of the quantity $T/n^{2/3}$ which
characterizes the entropy of the gas. Bottom panel shows the
distribution of tracer particles, originally located within the
bubble. The left column corresponds to the state shortly after the
beginning of simulations and the right column corresponds to the state
after $\sim 2.6 \times 10^8$ years. The initial entropy of the bubble
(green circle in the top-left plot) was such that the cluster gas has
a similar entropy at a radius of 25 kpc from the center. The entropy
contrast between the bubble and surrounding cluster gas is rather
modest and the bubble travels only a small distance from the center.
The contours of the initial bubble and a circle with a 25 kpc radius
are shown by black lines. The outer edge of the simulated volume (outer
black line in the bottom panel) corresponds to a distance of 50 kpc
from the cluster center. It is clear that at the end of the simulation,
tracer particles are concentrated along the surface where
cluster gas has a similar entropy to that of the gas in the original
bubble.
\label{pan}
}
\end{figure*}

A simple simulation demonstrating pancake formation is presented in
Fig.\ref{pan}. In this simulation, we assumed that the bubble is
initially filled with gas having the same entropy as the cluster gas
at a distance of $\sim$25 kpc. This is done by setting the temperature
of the gas inside the bubble a factor of $\sim$2 higher than the
temperature of the surrounding cluster gas. The density in the bubble
was set a factor of 2 lower in order to maintain pressure
equilibrium. The resulting modest contrast in entropy between the
bubble and the surrounding gas approximately models the situation when
the filling factors of the radio emitting plasma and thermal plasma in
the bubble are comparable as considered in the previous section. The
upper row of Fig.\ref{pan} shows the entropy distribution at the
beginning of the simulation and after $\sim 2.6 \times 10^8$ years. In
the lower panel, we show the distribution of tracer particles, which
initially are located within the bubble, at the same times. The two
black curves show, respectively, the boundary of the bubble at the
beginning of the simulation and the region in the cluster where the
cluster gas has the same entropy as the gas initially inside the
bubble. It is clear that at the end of the simulation, tracer
particles are concentrated along a surface where cluster gas has
a similar entropy to that of the gas in the original
bubble. Pancake--like structures are already visible after $\sim 10^8$
years from the start of the simulations. Oscillations near the neutral
buoyancy point were also observed.

Strongly underdense bubbles may rise to much larger distances from the
cluster center  before reaching a neutral buoyancy
point (see Fig.\ref{rfin}). We did not simulate the evolution of strongly
underdense bubbles to such late stages, but qualitatively
the picture should be the same.

The largest structures visible in Figure \ref{fig:m87} are the two
almost circular low surface brightness emission regions to the
northeast and southwest of the centre. Owen et al (2000) suggest that
these are spherical bubbles of radio plasma seen in projection. In this
case, they may receive substantial energy input from the inner
lobes which contain the active jet. Alternatively one can 
identify the circular emission regions seen in the radio images with
earlier bubbles that have reached their isodensity 
distance and are now in the process of spreading  and forming
pancake-like structures. We stress again that this requires that either relativistic
and thermal particles are mixed on microscopic scales or the lumps
of relativistic plasma and the ambient medium are bound together.

Owen et al. (2000) have noted the sharp boundary of the
M87 halo which is traced by all observations at different
frequencies and they argue that this excludes a dominant
diffusive transport process for the relativistic plasma.
In the proposed model the radius and sharpness of the
boundary is naturally explained by the height to which 
buoyant bubbles can rise. Also the complicated structure
of the outer halo may in this model reflect several layers
of partly fragmented bubbles.

The time required for pancake formation at large distances
from the cluster center is at least several times
longer than that required for a strongly buoyant bubble to travel such
distances. Therefore, it is extremely difficult for relativistic particles to
survive (see \ref{sec:sync}). Either reacceleration or exposure of
relativistic particles to stronger magnetic fields seems to be
inevitable to explain the radio emission from the outer radio halo if it
is indeed the result of lateral spreading of a bubble which has
reached the neutral buoyancy point.

If, on the contrary, lumps of radio emitting plasma are not bound to
the captured lumps of thermal gas, then radio bubbles will continue
to rise along the pressure gradient and will eventually leave the
central cooling flow region. Overdense lumps of  captured 
gas will separate from the bubbles and fall back towards the central
region. One possible example of overdense material having fallen back
toward the central galaxy may be found in A1795. Cowie et al. (1983)
reported a  remarkable $\sim45''$ (41 kpc long for the Hubble constant of
100 km s$^{-1}$ Mpc$^{-1}$) filamentary emission line (H$\alpha$ +
[NII])  chain originating near the central cD galaxy. 

\subsection{Total energy budget}

The X--ray luminosity of the whole cooling flow region around M87 is
about $10^{43}$ erg s$^{-1}$. On the other hand, estimates of the total
energy supplied by the jet to the ICM give a much larger value of
$\sim 10^{44}$ erg s$^{-1}$ (e.g. Owen, Eilek and Kassim 2000).  If all this
energy can be dissipated into the thermal gas in the cooling flow
region then the cooling flow should instead be a ``heating'' outflow
(Owen, Eilek and Kassim 2000), at least at the present moment. The
situation may be transient  so that
inflow and outflow periods alternate depending on the strength of
the AGN activity (e.g. Tucker and David 1997, Binney 1999, Owen, Eilek and
Kassim 2000). We speculate below on the consequences of a large input
of power in the form of relativistic plasma into the cooling flow
region.

Initially the radio lobe, inflated by the jet, expands supersonically
(e.g. Heinz, Reynolds, Begelman 1998). Later, as the size of the radio
lobe grows, the expansion becomes subsonic. As we discussed above
(Section \ref{sec:veloc}), a
large and strongly underdense bubble would rise with approximately the
Keplerian velocity of $\sim 400$ km s$^{-1}$. Assuming that this velocity limits the
growth of the original bubble due to the jet input power $L$, one can write:
$\frac{\Gamma}{\Gamma-1}P_04\pi r^2 v=L$, where $P_0$ is the ambient
gas pressure, $\Gamma=4/3$ is the adiabatic index of the relativistic
gas in the bubble, $r$ is the size of the bubble. Substituting $L\sim
10^{44}$ erg s$^{-1}$, $P_0\sim 2 \times 10^{-10}$ erg~cm$^{-3}$ and
$v \sim 400$ km s$^{-1}$
we find $r\sim 5$ kpc, which is approximately the size of the inner
lobes surrounding the jets. If the subsonic expansion
phase is longer than the duration of the initial strongly supersonic
expansion, then most of the energy deposited by the AGN will go into the
enthalpy of the expanded lobe and not into shock heating the
cluster gas.

The inner lobes break loose (during a time $t\sim r/v\sim 10^7$ years) and
form bubbles 
which then subsonically rise through the cluster atmosphere (Gull and
Northover 1973). 
At any given moment, thermal gas tries to settle in the
potential of the cluster in order to form a nondecreasing entropy
profile. Rising bubbles produce convective motions in the
gas and  mix lumps of thermal gas with different
entropies. Using tracer particles added to the thermal gas (see
Fig.\ref{uplift}) we
estimated that 
some 3--4$\times 10^7M_\odot$ of cool, thermal gas has
been uplifted from below 15 kpc to above 35 kpc 
by the bubble after
$\sim 8\times 10^7$ years. 
Much larger masses of
gas undergo smaller displacements by that time. For example, $\sim 1.3 \times
10^8M_\odot$ has been lifted from below 14 kpc to above 25
kpc. Although these are very crude estimates, which are sensitive to
the particular initial and boundary conditions adopted in our
simulations, one can expect that if the bubbles are launched
approximately every $10^7$ years then they are able to transport some
1--10 $M_\odot$ of cold gas per year from the center to the
periphery of the cooling flow. Interestingly this value is comparable
to the mass deposition rate in M87 estimated from the X--ray data.
One can further list several important consequences of the
convection induced by the rising bubbles:

\begin{itemize}
\item Due to uplifted cold gas, the X--ray surface brightness
distribution should be shallower than in the absence of bubbles. The
entropy profiles also flatten. 
\item If the bubbles are  formed and are rising systematically over
some preferred direction then the cold gas will be continuously uplifted
over this direction while it will be flowing inward in other directions.
\item Thermal instabilities can effectively take place within an
environment of extensive convective motions. As a result,
distributed mass deposition from cooling gas can naturally occur.
\item Overdense lumps of thermal gas, uplifted by the rising bubbles,
may fall back (when separated from the ``carrier'' buoyant bubble) to
the central region thereby producing filamentary structures, stretching  
radially from the galaxy.
\end{itemize}

Even if the bubbles are not mixed on microscopic scales with the thermal
plasma and are able to eventually leave the cooling flow region, most
of the energy has to be deposited in the thermal gas (if there is a
strong contrast in pressure between the inner and outer regions of the
cooling flow). For a bubble moving at a constant terminal velocity
the dissipation of energy is characterized by the change of the bubble
enthalpy $\frac{\Gamma}{\Gamma-1}PV\propto
(P/P_0)^{\frac{\Gamma-1}{\Gamma}}$, where $P$ 
is the pressure of the cluster thermal gas.  The bubble energy 
goes into sound waves, internal waves, turbulent motion in the
wake, potential energy of the uplifted cold gas etc. Sound waves can
carry energy away from the central region, while most of the other
channels would eventually result in heating the cooling flow region. 
If the jet power is indeed $\sim 10^{44}$ erg s$^{-1}$, then even a modest
10\% efficiency of local dissipation into heat should be enough to
exceed the radiative cooling of the gas. The gas then heats up, 
but this heating is distributed through a large
volume. Since no shock waves are envisaged in this picture one would
not expect to find localized, hot, compressed regions. 

If strong energy input from AGN is maintained for a very long period
(and dissipation into heat is efficient), then a core with a flat
entropy profile will be formed and the size of this core will grow
with time. We note here that the rate of energy extraction from
the bubble depends on the pressure gradient (see above expression for
enthalpy). Therefore, if the gas is heated to a temperature comparable
to the depth of the local potential well, the dissipation efficiency
will decrease in this region. This effect may provide a
self-regulation for heating in the central part of the cooling flow
region.

Eventually, strong heating is capable of completely eliminating the
cooling flow, unless the power of 
an AGN is in turn regulated by the conditions in the cooling flow region as
discussed by Tucker and David (1997), Binney (1999), Owen, Eilek and
Kassim (2000). In M87, heating at a rate of $\sim 5 \times 10^{43}$
ergs s$^{-1}$
would eliminate the cooling flow for $\sim 10^8$ years. This is
comparable to the age of the largest radio structures. On the
other hand, the complicated radio halo morphology suggests
that periods of high AGN activity have alternated with the periods of
low AGN power. We note here that dissipation of the subsonic motions
into heat may take a longer time than the characteristic 
bubble rise time and therefore the actual gas heating rate may be
related to the AGN activity averaged over a long period of time.

\section{Conclusions}

Buoyant bubbles of cosmic rays, slowly rising through the cooling gas,
can qualitatively explain the complicated radio and X--ray morphology
of the central 40 kpc region around M87. Torus--like features, seen in
the radio, may be similar to the ``mushrooms'' which appear in 
Rayleigh--Taylor unstable configurations: as the fluid rises through the
ambient medium, Kelvin-Helmholtz instabilities create torus-like
``mushroom'' heads. The excess X--ray emission
trailing some prominent radio features could be due to cold gas
captured by the rising bubbles and uplifted to large distances from
the central source. 

Rising bubbles produce strong convection in the cooling flow
region. Convection flattens X--ray surface brightness and
entropy profiles in the cooling flow. Convection may promote
distributed mass deposition in the cooling flow and the formation of the
filaments. If more than 10\% of the energy supplied by the jet is
dissipated into heat in the cooling flow region, then large
cores with flat entropy profiles may be formed.

New Chandra and XMM observations of cooling flow clusters
(e.g. McNamara et al. 2000, Fabian et al. 2000a) demonstrate that the
interaction of the radio sources and thermal gas is widespread in these
objects. Furthermore, the mass of cooling gas at temperatures less than
$\sim$1 keV, derived from the spectra, seems to be significantly
smaller than expected from the simplest cooling flow models (B\"ohringer et
al. 2001, Tamura et al. 2001, Peterson et al. 2001). Possible 
explanations for this discrepancy are summarized by Fabian et
al. (2000b; see also references therein). As we discussed above, heating by the
buoyant bubbles may play an important role in the thermal balance in
cooling flows. One of the predictions of the buoyant bubble picture,
namely a correlated morphology of the colder uplifted gas and radio
emission, seems to be in a broad agreement with recent XMM results
(Belsole et al. 2001).

Our model is of course highly oversimplified and is not intended to
closely reproduce the M87 environment. Clearly more detailed
simulations are needed, perhaps in 3D, which would also follow the
formation of the initial bubble (possibly with continuous energy input)
and include radiative cooling. However, despite the simple treatment
of bubbles in cluster and galaxy atmospheres presented above, we
believe that this approach indicates areas of particular interest.

\acknowledgments

We are grateful to the editor Steven N. Shore and the referee for 
important comments and suggestions. We thank Alexei Kritsuk, Nail
Inogamov, Christine Jones, Nail Sibgatullin and Henk Spruit for useful
discussions. This research has made use of data obtained through the High 
Energy Astrophysics Science Archive Research Center Online Service, provided 
by the NASA/Goddard Space Flight Center.

\label{lastpage}

\end{document}